# Towards inclusive practices with indigenous knowledge

Astronomy across world cultures is rooted in indigenous knowledge. We share models of partnering with indigenous communities involving collaboration with integrity to co-create an inclusive scientific enterprise on Earth and in space.

Aparna Venkatesan, David Begay, Adam J. Burgasser, Isabel Hawkins, Ka'iu Kimura, Nancy Maryboy and Laura Peticolas

Astronomy is recognized as the oldest human science, connected worldwide for millennia by celestial cycles, objects, and events recorded in the calendars and rituals of our ancestors. Astronomy therefore has a unique grounding in the cultural and scientific practices of indigenous peoples through indigenous knowledge (IK). We use 'indigenous' as a general term designating the original inhabitants of any land, who existed or exist for thousands of years in an explicitly interconnected local system of relationships including community, environment, and sky[1,2]. By IK, we mean a living, language-rooted practice unique to each indigenous culture and adapted to their specific environment, framed through the lens of their cultural values and spiritual traditions. IK is inherently interdisciplinary, multigenerational, and expressed through sustainable practices[3]. Additionally, connection to place has been identified by numerous researchers, including traditional knowledge holders, as an essential component in indigenous education[4,5].

Astronomy's deep grounding in IK translates to an equally strong obligation for modern astronomical institutions to partner with indigenous communities, especially when indigenous lands or resources are involved. This is an issue across nearly all continents at present, given the role of colonization and the related global aftermath, and is poised to become a cosmic issue given the pace and manner in which we are occupying near-Earth and interplanetary space. With a greater awareness of IK and indigenous perspectives, nations and governing bodies have an opportunity to learn practices that lead to sustained healthy communities. The possibility to move towards global consensus-building also increases with IK's approach to conflict resolution and interconnections. This is especially relevant as most of us are schooled in Western models of science and scientific achievement.

As we face unprecedented global challenges and crises nearly two decades into this century, we are called on to move beyond mere diversity/equity considerations, for example, the longstanding underrepresentation of indigenous people in science, technology, engineering and mathematics (STEM) fields within institutions or academia. Astronomy must be prepared to do the patient long-term work of deep listening, genuine dialogue with indigenous communities and leaders, and recognition of the great scientific value of the multigenerational data gathering and experiential wealth contained holistically in IK. This approach has the potential for innovative solutions from all human ways of knowing, and could surprise us with progress that is richer than from one perspective alone[6].

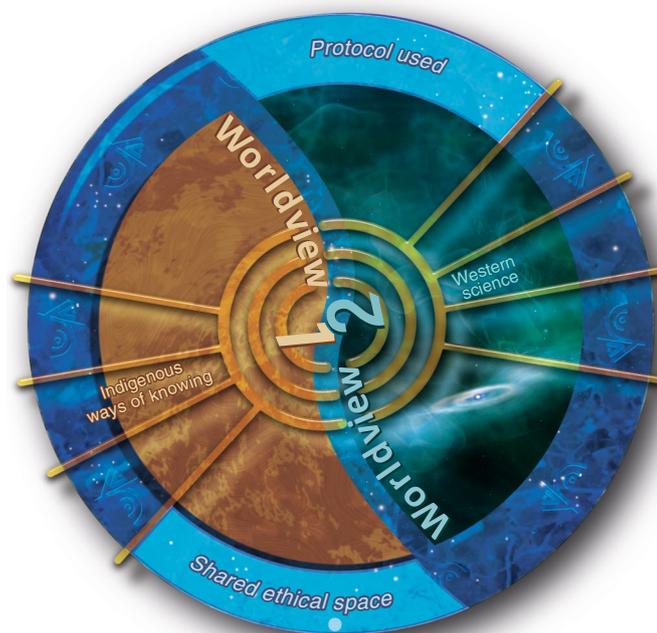

**Fig. 1 | Representation of what is required to co-create in collaboration with people with differing worldviews.** The outer boundary of the circle is a boundary of ethics and protocol. This boundary is a one-way membrane at its outer edge: only ethical and positive protocols meant to respect and support each worldview comes into the collaboration space. When entering this space, negative expectations, prescribed views, or unconstructive attitudes are left outside the circle. This is a 2D representation of a 3D space to allow for the worldviews to overlap in some cases. Credit: Aurore Simonnet, Sonoma State University / Education & Public Outreach

We share here a number of successful practices and programmes that build on the strength and potential of IK in collaboration with astronomy, drawing from our collective experience as professional astronomers and knowledge holders in a variety of formal and informal astronomy research and learning environments.

### Successful collaborations with IK

In a recent white paper (https://arxiv.org/abs/1908.02822) submitted to the Decadal Survey on Astronomy and Astrophysics (Astro2020) in the category of Activities, Projects, or State of the Profession Consideration (APC), we shared six key recommendations for US funding agencies and institutions that are important first steps for nonindigenous institutions to fully partner with indigenous communities and IK[7]. These include: collaboration pathways for telescopes and other research facilities on indigenous lands; dedicated funding for successful models that coherently synergize indigenous and nonindigenous science; dedicated funding of long-term, observation-based interdisciplinary research conducted by indigenous knowledge holders; recognizing oral traditions as IK's equivalent of peer-reviewed, peer-verified scientific results; dedicated conference funding for indigenous speakers and leaders; and a culturally supported path for full participation of indigenous youth in science careers. Collectively, these recommendations honour the intergenerational, interdisciplinary scientific research underlying IK.

A number of highly successful initiatives serve as roadmaps for how to integrate indigenous and nonindigenous knowledge systems. This model of partnerships, called Collaboration with Integrity[3], was pioneered by the Indigenous Education Institute, the 'Imiloa Astronomy Center of Hawai'i, and I-WISE (Indigenous Worldviews in Informal Science Education), and includes: Cosmic Serpent, A Hua He Inoa, EnVision Maunakea, Native Universe, and Maunakea Scholars. These initiatives demonstrate the co-creation of educational and professional development experiences that involve understanding and embracing the deep historical, cultural, and traditional language base of IK. Such collaboration is possible by avoiding the historical pattern of appropriation or assimilation into nonindigenous knowledge systems[8]. As members of the astronomical community, we must challenge ourselves to have the imagination to go beyond divisive dualities.

At the 2015 I-WISE conference, the Collaboration strand workshop participants built consensus around what is needed to collaborate with integrity. It was agreed that such collaboration has a high probability of success in any situation when a shared ethical space and specific protocols for communication are developed before attempting to partner. Figure 1 provides a graphical description of what is needed for two communities with different worldviews to work together for the mutual benefit of both communities. For example, it is recommended that astronomers and the local indigenous community living on or near a mountain develop ethical and transparent protocols for communication prior to establishing scientific motivations and telescope or facility designs. This can be applied to both existing and new telescopes on indigenous lands, and to the rapidly accelerating human presence in space, both of which are discussed in more detail below. The workshop participants developed a detailed series of recommendations in order to collaborate with integrity (see the I-WISE conference site), including: initiating in-person dialogue with community leaders; engaging all stakeholders at the beginning of an initiative; creating safe spaces and diverse communication pathways; co-creating goals as equals from the start; sharing authority; understanding each other's privilege; and funding capacity rather than projects.

Additionally, the I-WISE Next Generation strand workshops recognized the key role that youth, in particular Native youth, play in making IK visible and explicit in STEM fields and careers. There were sixteen participants from 17 to 30 years of age, representing 14 different indigenous cultures from the continental United States, Hawai'i, and Mexico (some of these participants are shown in Fig. 2). These participants shared the numerous ways in which Western institutions would be different if they included indigenous perspectives, including long-term thinking, sustainability, experiential learning, community-oriented values instead of individualistic and capitalistic values, more autonomy over ancestral sites, and shared knowledge between elders and youth.

### Telescopes on indigenous lands

Astronomical facilities can create meaningful pathways for transformative collaboration and research opportunities to work with local indigenous communities and youth education/research programmes. One approach is to empower such communities to name objects discovered with telescopes on indigenous lands. The power of naming is a belief common to nearly all indigenous peoples worldwide (see, for example, ref.9), as IK is deeply rooted in the traditional language base of each indigenous community. Examples of collaborative naming by nonindigenous and Native Hawaiian scholars include the interstellar asteroid 'Oumuamua[10], and Pōwehi[11], the first imaged black hole.

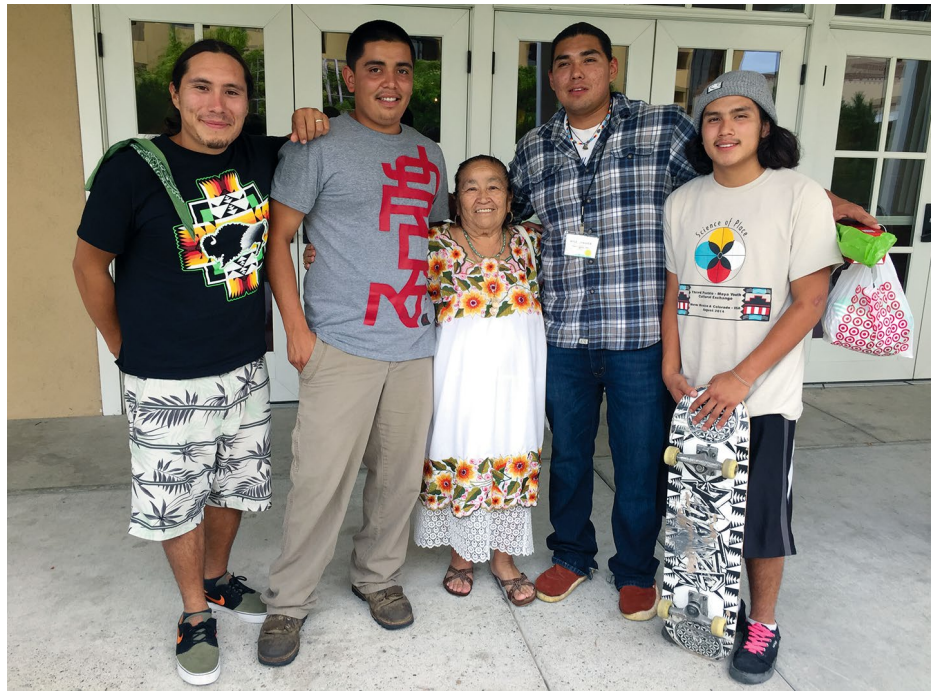

**Fig. 2 | Conference participants at the 2015 I-WISE meeting in Albuquerque, New Mexico, USA.** From left to right: Angelo McHorse, Emilio Torivio, Maria Avila Vera, Kyle Swimmer, and Victor Corpuz.

The A Hua He Inoa (AHHI) nomenclature project, developed by the 'Imiloa Astronomy Center of Hawai'i in partnership with Hawai'i-based astronomical observatories, is doing groundbreaking work to weave traditional indigenous practices into the official naming of astronomical discoveries made in Hawai'i. This collaborative effort integrates Native Hawaiian-speaking youth into the process of naming astronomical discoveries, promotes youth engagement with astronomers and Hawaiian language experts, and allows all to gain a deeper understanding of Hawaiian and scientific knowledge on the origins of the Universe. Additionally, AHHI creates pathways in which language and culture are at the core of modern scientific practice. Efforts like AHHI broaden the cultural impact of progress in astronomy while developing new contexts for IK and language that ensure the vitality of indigenous communities and their identities.

We end this section by respectfully acknowledging the presence and practices of the protectors, activists, and scientists on Maunakea. We honour this moment as an opportunity for greater awareness and intentional partnerships between IK and mainstream astronomy.

## Communities, not colonies, in space

Space is becoming the playground of billionaire investors and private corporations. In the past decade alone, the role of the private sector in space has grown dramatically in scope and scale, including satellite launches, asteroid mining, space tourism, and interplanetary exploration. The policies of colonialism/imperialism have had a strong impact on indigenous communities worldwide and the preservation of IK practices. Without consistent international regulatory and ethical standards, we are in danger of expanding such policies rooted in the mindset of colonialism to a truly cosmic scale.

In our Astro2020 APC white paper, we advocated for US federal funding agencies to "incentivize public and private organizations and institutions, including professional societies, to include the perspectives of indigenous astronomers and scientists in the rapid acceleration of human activity and presence in space." This was a key theme emphasized in another Astro2020 APC white paper (https://arxiv.org/abs/1907.05834) on the future of space exploration and science missions[12]. These authors stressed the improvement of current standards for safety, ethics, biocontamination, and planetary protection. They also advocated for "moving international space law and domestic space policy from a reactive nature to a proactive one [that] will ensure the future of space exploration is one that is safe, transparent, and anti-imperialist", and for "anti-colonization standards and protocol to ensure equal and fair participation in space". These authors also call on the astronomical community to prevent further ground-based colonialism, and to factor in ecological, environmental, and indigenous concerns when building new astronomical facilities.

The core issue is how we formulate who space belongs to. Do we treat it as a shared resource held in community trust, much like national parks or clean air and water; or do we accept the first-come, first-claim strategy that is already occurring in a race to ownership? We advocate for dark skies being a human right, and call on all nations to recognize the importance of dark skies to the cultural practices of indigenous peoples, especially for celestial navigators and the practice of wayfinding by those cultures of Polynesian descent (Hawai'i, Māori, and Oceania peoples). The successful launch of dozens of small reflective satellites from SpaceX, and many more planned by Amazon, create a night sky that could soon have tens of thousands of modest-albedo satellites, well exceeding the number of stars visible to the human eye[13]. These systems undermine many ground-based astronomical observing projects as well as the efforts to preserve dark skies in ever-decreasing areas of our planet. Although such corporate launches promote global equality through internet satellite constellations (for example, SpaceX's Starlink), the decisions to create these systems must be carefully and collaboratively reached through international consensus including indigenous leaders and representatives. The outcomes will reverberate for all of humanity, not just for astronomy (see the position statement by the American Astronomical Society).

## Looking ahead

Science, like our planet, is facing dwindling resources to address a large array of issues; solutions must move us towards a more sustainable and inclusive future. The path to sustainable solutions is fragile and challenging, with consequences for many generations to come. This complex situation is a wake-up call for greater awareness of IK and greater impetus to build true partnerships that integrate IK and mainstream astronomy. It is also a dynamic opportunity for co-creating a more inclusive scientific endeavour with intent and respect. IK is a lived practice demonstrating how science is inseparable from the people conducting the science. Most of us are used to leaving behind our cultural beliefs and identities at the door when conducting our scientific work. Rather than presenting an inauthentic choice between science and culture to indigenous youth and future scientists, we can have the collective imagination and resolve to transform scientific culture with IK. As in a healthy ecology, we can integrate IK practices and principles to catalyse the creative solutions and innovations that are much needed for the future. Please contact any of the indigenous centres and initiatives mentioned here for information and resources on how to include IK and indigenous perspectives in schools and institutions. ❐

Aparna Venkatesan[1]*, David Begay[2,3], Adam J. Burgasser[4], Isabel Hawkins[5], Ka'iu Kimura[6], Nancy Maryboy[2,3] and Laura Peticolas[7]

[1]University of San Francisco, San Francisco, CA, USA. [2]Indigenous Education Institute, Friday Harbor, WA, USA. [3]University of Washington, Seattle, WA, USA. [4]University of California San Diego, La Jolla, CA, USA. [5]San Francisco Exploratorium, San Francisco, CA, USA. [6]'Imiloa Astronomy Center of Hawai'i, Hilo, HI, USA. [7]Sonoma State University, Rohnert Park, CA, USA.
*e-mail: avenkatesan@usfca.edu

### Acknowledgements
A.V. gratefully acknowledges support from the University of San Francisco Faculty Development Fund. We dedicate this Comment to the world's wealth of indigenous knowledge and to two beloved indigenous colleagues who were taken too soon: Dr 'Auntie' Verlie Ann Malina-Wright, Hawaiian–Chinese–Irish educator of over 50 years and wise elder; and Dr Paul Coleman, cherished field mentor and the first Native Hawaiian to earn a doctorate in astrophysics. As Paul said to his family, "I offer the lesson of the stone mason; the greatest works require a tremendous effort with surprising patience, one stone at a time."